\documentstyle[12pt]{article}

\def \vslash{\!\!\not{\! v}}

\begin{document}

\renewcommand{\thefootnote}{\fnsymbol{footnote}}

\begin{titlepage}

\begin{flushright}
IJS-TP-94/19\\
TUM-31-62/94 \\
NUHEP-TH-94-9\\
July 1994\\
\end{flushright}

\vspace{.5cm}

\begin{center}
{\Large \bf  Vector and pseudoscalar charm meson radiative decays\\}

\vspace{1.5cm}

{\large \bf B. Bajc $^{a}$, S. Fajfer $^{a,b}$ and
Robert J. Oakes $^{c}$\\}

\vspace{.5cm}

{\it a) Physics Department, Institute "J.Stefan ", Jamova 19,
61111 Ljubljana, Slovenia\\}

\vspace{.5cm}

{\it b) Physik Department, Technische Universit\"at M\"unchen,
85748 Garching,
FRG}

\vspace{.5cm}

{\it c) Department of Physics and Astronomy
Northwestern University, Evanston, Il 60208
U.S.A.}

\vspace{1cm}

\end{center}

\centerline{\large \bf ABSTRACT}

\vspace{0.5cm}

Combining heavy quark effective theory
and the chiral Lagrangian approach we investigate
radiative decays of pseudoscalar $D$
mesons. We first reanalyse $D^{*} \rightarrow D \gamma$
decays within the effective Lagrangian approach
using heavy quark spin symmetry, chiral symmetry
Lagrangian, but including also the light vector
mesons. We then investigate $D \rightarrow V \gamma$
decays and calculate the $D^0 \rightarrow \bar{K}^{*0} \gamma$
and $D^{s+} \rightarrow \rho^+ \gamma$ partial widths and
branching ratios.
\end{titlepage}

\setlength {\baselineskip}{0.75truecm}
\parindent=3pt  % no indention of new paragraphs
\setcounter{footnote}{1}    % start footnotes at dagger instead
			     % of '*' (article style only)

\newcommand{\tr}{\mbox{\rm Tr\space}}
\def \vslash{\!\!\not{\! v}}
\renewcommand{\thefootnote}{\arabic{footnote}}
\setcounter{footnote}{0}

\vspace{.5cm}

\begin{center}
{\bf 1.Introduction}
\end{center}

The list of $D$ meson decay rates is rather
long and further study of their decays would
eventually help to better understand their
features. There has not yet been any experimental
evidence for radiative decays of $D$ mesons,
while $D^{*}$ radiative decays are known
to be important. The $D^{*}$ decays \cite{CG,ABJ}
can be described in a model independent
framework which incorporates the appropriate constraints
on the decay amplitudes. The combination of
heavy quark effective theory and chiral Lagrangians
have been extensively studied and applied to
many $D$ mesons decays
\cite{IW,BD,YCC,LLY,PC,HG1,HG2,FGL,G1,G2,G3}.
Wise \cite{MW} has
proposed an effective Lagrangian to describe,
at low momentum, the interactions of a  meson
containing a heavy quark with the light
pseudoscalar mesons $\pi$, $K$, $\eta$.
Two kinds of symmetries characterize the effective
Lagrangian: the heavy quark $SU(2)$ spin symmetry and
the non-linearly realized $SU(3) \otimes SU(3)$
chiral symmetry in the light sector, corresponding
to spontaneous symmetry breaking of the chiral
group to the diagonal $SU(3)_{V}$. Due to the
rather large masses of the $D$ mesons, the inclusion
of resonances with masses below the $D$ mesons
seems necesssary \cite{G1,G2,G3}.

In this paper, following the requirements of heavy
quark and chiral symmetry, we develop a framework for
the description of heavy and light pseudoscalar and
vector mesons. In section 2 we write down the most
general Lagrangian in the limit of exact heavy quark
and chiral symmetries. Section 3 is devoted to the higher
order odd parity Lagrangian, which also describes the
decay $D^* \to D \gamma$ , and we
reinvestigate this decay in order to learn
more about the couplings in the chiral Lagrangian. In section
4 we analyse the weak Lagrangian. Finally, as an example of
the use of our model, we calculate the $D \to V \gamma$
radiative decays in section 5.

\vspace{.5cm}

\begin{center}
{\bf 2. The chiral Lagrangian technique and heavy quark limit}\\
\end{center}

The strong interaction meson Lagrangian for the light
pseudoscalar octet and heavy pseudoscalar
and vector triplets in the chiral and heavy
quark limits
was first written down by Wise \cite{MW}
(see also \cite{BD}). The electromagnetic
interactions between these
mesons was described in \cite{CG,ABJ,LLY}.
The octet of light vector mesons
was included in the Wise Lagrangian
\cite{MW} later by Casalbuoni, et al.
\cite{G2} as the gauge particles associated
with the hidden symmetry group SU(3)$_H$
\cite{BKY}. The next step is to provide a
common description of both the light and
heavy pseudoscalar mesons, which also includes
both light and heavy vector mesons and the
electromagnetic
interactions. In this section we present
the strong and electromagnetic Lagrangian for
the description of both light and heavy
pseudoscalar and vector mesons.

The light pseudoscalar mesons are described by the
$3 \times 3$ unitary matrix

\begin{eqnarray}
\label{eone}
u & = & \exp  ( \frac{i \Pi}{f} )
\end{eqnarray}

where $f \simeq 132$ MeV is the pion pseudoscalar
pion constant
and $\Pi$ is the pseudoscalar meson
unitary matrix defined as

\begin{eqnarray}
\label{etwo}
\Pi = \pmatrix{
{\pi^0 \over \sqrt{2}} + {\eta \over \sqrt{6}} & \pi^+ & K^+ \cr
\pi^- & {-\pi^0 \over \sqrt{2}} + {\eta \over \sqrt{6}} & K^0 \cr
K^- & {\bar K^0} & -{2 \over \sqrt{6}}\eta \cr}
\end{eqnarray}

The octet of light vector mesons is described
by the $3 \times 3$ unitary matrix

\begin{eqnarray}
\label{ethree}
{\hat \rho}_\mu & = & i {g_V \over \sqrt{2}} \rho_\mu
\end{eqnarray}

where $g_V$ ($\simeq 5.8 \sqrt{2/a}$ with $a=2$ in
the case of exact vector dominance) is the coupling
constant of the vector meson self--interaction
\cite{BKY} and $\rho_\mu$ the vector meson
unitary matrix

\begin{eqnarray}
\label{efour}
\rho_\mu = \pmatrix{
{\rho^0_\mu + \omega_\mu \over \sqrt{2}} & \rho^+_\mu & K^{*+}_\mu \cr
\rho^-_\mu & {-\rho^0_\mu + \omega_\mu \over \sqrt{2}} & K^{*0}_\mu \cr
K^{*-}_\mu & {\bar K^{*0}}_\mu & \Phi_\mu \cr}
\end{eqnarray}

In the following we will also use the gauge field
tensor $F_{\mu \nu} ({\hat \rho})$, defined as

\begin{eqnarray}
\label{efive}
F_{\mu \nu} ({\hat \rho}) & = &
\partial_\mu {\hat \rho}_\nu -
\partial_\nu {\hat \rho}_\mu +
[{\hat \rho}_\mu,{\hat \rho}_\nu]
\end{eqnarray}

The heavy mesons are $Q{\bar q}^{a}$ ground states,
where $Q$ is a c (or b) quark and
$q^{1} = u$, $q^{2} = d$ and $q^{3} = s$.
In the heavy quark limit they are
described by $4 \times 4$
matrix $H_{a}$ $(a = 1,2,3)$ \cite{MW}

\begin{eqnarray}
\label{esix}
H_a& = & \frac{1}{2} (1 + \vslash) (P_{a\mu}^{*}
\gamma^{\mu} - P_{a} \gamma_{5})
\end{eqnarray}

where $P_{a\mu}^{*}$ and $P_{a}$ annihilate,
respectively, a spin-one and spin-zero meson
$Q{\bar q}^{a}$ of velocity $v_{\mu}$. The
creation operators $P_{a\mu}^{* \dag}$ and $P_{a}^{\dag}$
occur in \cite{MW}

\begin{eqnarray}
\label{eseven}
{\bar H}_{a} & = & \gamma^{0} H_{a}^{\dag} \gamma^{0} =
(P_{a\mu}^{* \dag} \gamma^{\mu} + P_{a}^{\dag} \gamma_{5})
\frac{1}{2} (1 + \vslash)
\end{eqnarray}

Following the analogy in refs. \cite{BKY,G2} we
introduce two currents:

\begin{eqnarray}
\label{eeight}
{\cal V}_{\mu} & = & \frac{1}{2} (u^{\dag}
D_{\mu} u + u D_{\mu}u^{\dag})
\end{eqnarray}

and

\begin{eqnarray}
\label{enine}
{\cal A}_{\mu} & = & \frac{1}{2} (u^{\dag}
D_{\mu} u - u D_{\mu}u^{\dag})
\end{eqnarray}

where the covariant derivatives of $u$ and $u^{\dag}$ are
defined as

\begin{eqnarray}
\label{eten}
D_\mu u & = & (\partial_\mu + {\hat B}_\mu) u
\end{eqnarray}

and

\begin{eqnarray}
\label{eeleven}
D_\mu u^{\dag} & = & (\partial_\mu + {\hat B}_\mu) u^{\dag}
\end{eqnarray}

while

\begin{eqnarray}
\label{etwelve}
{\hat B}_\mu & = & i e B_\mu Q
\end{eqnarray}

\begin{eqnarray}
\label{ethirteen}
Q = \pmatrix{
{2 \over 3} & 0 & 0 \cr
0 & -{1 \over 3} & 0 \cr
0 & 0 & -{1 \over 3} \cr}
\end{eqnarray}

and $B_\mu$ is the photon field. To insure that the
vertices $D^{0 \dag} D^0 \gamma$ and
$D^{*0 \dag} D^{*0} \gamma$, or those with
D replaced by B in the case of the b quark,
are absent, we define the covariant
derivative for the heavy meson field as

\begin{eqnarray}
\label{efourteen}
D_\mu {\bar H}_a & = & (\partial_\mu +
{\cal V}_\mu - i e Q' B_\mu) {\bar H}_a
\end{eqnarray}

with $Q'=2/3$ for c quark ($-1/3$ for b quark).
With these definitions we can
finally write down the even parity
strong and electromagnetic
Lagrangian for heavy and light pseudoscalar and
vector mesons:

\begin{eqnarray}
\label{efifteen}
{\cal L}_{even} & = & {\cal L}_{light} -
{1 \over 4} (\partial_\mu B_\nu - \partial_\nu B_\mu)^2 +
i Tr (H_{a} v_{\mu} D^{\mu} {\bar H}_{a})\nonumber\\
& + &i g Tr [H_{b} \gamma_{\mu} \gamma_{5}
({\cal A}^{\mu})_{ba} {\bar H}_{a}]
 +  i \beta Tr [H_{b} v_{\mu} ({\cal V}^{\mu}
- {\hat \rho}_{\mu})_{ba} {\bar H}_{a}]\\
& + &  {\beta^2 \over 2 f^2 a}
Tr ({\bar H}_b H_a {\bar H}_a H_b)\nonumber
\end{eqnarray}

with

\begin{eqnarray}
\label{esixteen}
{\cal L}_{light} & = & -{f^2 \over 2}
\{tr({\cal A}_\mu {\cal A}^\mu) +
a\, tr[({\cal V}_\mu - {\hat \rho}_\mu)^2]\}\nonumber\\
& + & {1 \over 2 g_V^2} tr[F_{\mu \nu}({\hat \rho})
F^{\mu \nu}({\hat \rho})]
\end{eqnarray}

This Lagrangian is invariant under the following
gauge transformation:

\begin{eqnarray}
\label{eseventeen}
H & \to & e^{i e Q' \lambda(x)} H g_0^{\dag}(x)\nonumber\\
{\bar H} & \to & g_0(x) {\bar H} e^{-i e Q' \lambda(x)}\nonumber\\
u & \to & g_0(x) u g_0^{\dag}(x)\nonumber\\
u^{\dag} & \to & g_0(x) u^{\dag} g_0^{\dag}(x)\nonumber\\
{\cal V}_\mu & \to & g_0(x){\cal V}_\mu g_0^{\dag}(x)+
g_0(x)\partial_\mu g_0^{\dag}(x)\\
{\cal A}_\mu & \to & g_0(x){\cal A}_\mu g_0^{\dag}(x)\nonumber\\
{\hat \rho}_\mu & \to & g_0(x){\hat \rho}_\mu g_0^{\dag}(x)+
g_0(x)\partial_\mu g_0^{\dag}(x)\nonumber\\
{\hat B}_\mu & \to & g_0(x){\hat B}_\mu g_0^{\dag}(x)+
g_0(x)\partial_\mu g_0^{\dag}(x)\nonumber
\end{eqnarray}

where $g_0(x)=exp(i e Q \lambda(x))$. The last
transformation (\ref{eseventeen}) together with
(\ref{etwelve})-(\ref{ethirteen}) imply, of course, the
usual gauge transformation for the photon
field:

\begin{eqnarray}
\label{eeighteen}
B_\mu & \to & B_\mu - \partial_\mu \lambda(x)
\end{eqnarray}

In equation (\ref{efifteen}) $g$ and $\beta$ are
constants which should be determined from
experimental data \cite{CG,ABJ,G1,G2,G3}.
The constant $a$ in (\ref{efifteen})-(\ref{esixteen})
is in principle a free parameter, but we shall
fix it by assuming exact
vector dominance \cite{BKY}, for which
$a=2$. With exact vector dominance
there are no direct vertices between the
photon and two pseudoscalar mesons, so that the
pseudoscalars interact with the photon only
through vector mesons.

The electromagnetic field can couple to the
mesons also through the anomalous interaction;
i.e., through the odd parity Lagrangian. Even
with the $PP\gamma$ direct vertices absent
in ${\cal L}_{light}$ due to the choice $a=2$,
direct $PV\gamma$ vertices are present in
the odd parity Lagrangian. We write down the two
contributions which are significant for our
calculation:

\begin{eqnarray}
\label{enineteen}
{\cal L}^{(1)}_{odd} & = & -4 \frac{C_{VV\Pi}}{f} \epsilon
^{\mu \nu \alpha \beta}Tr (\partial_{\mu}
{\rho}_{\nu} \partial_{\alpha}{\rho}_{\beta} \Pi)
\end{eqnarray}

\begin{eqnarray}
\label{etwenty}
{\cal L}^{(2)}_{odd} & = & -4 e {\sqrt 2}
\frac{C_{V\pi\gamma}}{f} \epsilon
^{\mu \nu \alpha \beta}
Tr (\{ \partial_{\mu}{\rho}_{\nu},
\Pi \} Q \partial_{\alpha} B_{\beta})
\end{eqnarray}

Equation (\ref{enineteen}), together with vector dominance
couplings

\begin{eqnarray}
\label{etwentyone}
{\cal L}_{V-\gamma} & = & - m_V^2 {e \over g_V} B_{\mu}
(\rho^{0\mu} + {1 \over 3} \omega^{\mu} -
{\sqrt{2} \over 3} \Phi^{\mu})
\end{eqnarray}

which come from the second term in (\ref{esixteen}),
describe the electromagnetic
interaction assuming
vector-meson dominance, while the direct
photon-light vector meson-pseudoscalar
interactions are contained in (\ref{etwenty}).
The contributions to the odd Lagrangian (\ref{enineteen})
and (\ref{etwenty}) arise from
Lagrangians of the Wess-Zumino-Witten
kind \cite{WZ,EW}.

In the $m_q \to 0$ and $m_Q \to \infty$
limit ($m_q$ and $m_Q$ are the masses of the
light and heavy quarks, respectively) the
strong and electomagnetic interactions of
heavy and light pseudoscalar and vector
mesons are thus described by the even
Lagrangian (\ref{efifteen})-(\ref{esixteen}) and
by the odd Lagrangian (\ref{enineteen})-(\ref{etwenty}).
However the $D^* D \gamma$ vertices are not
included in the above Lagrangian
since it is of the anomalous type. The
terms responsible for it are of higher order
$1/m_Q$ and they will be introduced in the next section.

\vspace{.5cm}

\begin{center}
{\bf 3. Higher order odd Lagrangian for heavy mesons}\\
\end{center}

In our approach vector meson dominance
describes the couplings of light quarks
and photons through the higher dimensional
invariant operator

\begin{eqnarray}
\label{etwentytwo}
{\cal L}_{1} & = & i {\lambda} Tr [H_{a}\sigma_{\mu \nu}
F^{\mu \nu} (\hat \rho)_{ab} {\bar H_{b}}]
\end{eqnarray}

In this term the interactions of light
vector mesons, heavy pseudoscalars or
heavy vector $D$ mesons are also present. The
light vector meson can then couple to the photon
by the standard vector-meson dominance prescription
(\ref{etwentyone}).
These terms effectively describe the light quark-
photon interaction inside the charmed (or beauty)  mesons.

The coupling $\lambda$ can be independently
determined either from $D^{*}$
decays into $D \pi$ and $D \gamma$
\cite{CG,ABJ,YCC,LLY}, some ratios of which
have been measured \cite{PD,expr}
or from semileptonic decays of $D$
mesons \cite{G1,G2,G3}.

The heavy quark-photon
interaction is genereted by the term

\begin{eqnarray}
\label{etwentythree}
{\cal L}_{2} & = & - \lambda^{\prime}
e Tr [H_{a}\sigma_{\mu \nu}
F^{\mu \nu} (B) {\bar H_{a}}]
\end{eqnarray}

According to quark models
the parameter$|\lambda^{\prime}|$ can be
approximately related to
the charm quark magnetic moment via
$1/(6 m_c)$ \cite{CG,ABJ,YCC,LLY}.
In order to reduce the error in determinig
the couplings we shall reanalyze the decays
$D^{*}\rightarrow D \pi$
and $D^{*}\rightarrow D \gamma$. Experimentally
one measures \cite{expr} the branching fractions
$R_{\gamma}^0 = \Gamma (D^{*0} \to D^0 \gamma) /
\Gamma (D^{*0} \to D^0 \pi^0)=0.572 \pm 0.057 \pm 0.081$.
and $R_{\gamma}^+ = \Gamma (D^{*+} \to D^+ \gamma) /
\Gamma (D^{*+} \to D^+ \pi^0)=0.035 \pm 0.047 \pm 0.052$.
Using our Lagrangian these branching ratios are

\begin{eqnarray}
\label{etwentyfour}
R_{\gamma}^0 & = & 64 \pi f^2 \alpha_{EM}
({\lambda' \over g} + {2 \over 3} {\lambda \over g} )^2
({p_{\gamma}^0 \over p_{\pi}^0})^3
\end{eqnarray}

\begin{eqnarray}
\label{etwentyfive}
R_{\gamma}^+ & = & 64 \pi f^2 \alpha_{EM}
({\lambda' \over g} - {1 \over 3} {\lambda \over g} )^2
({p_{\gamma}^+ \over p_{\pi}^+})^3
\end{eqnarray}

To determine $\lambda/g$ and $\lambda'/g$,
the square-roots of the left--hand--sides of eq.
(\ref{etwentyfour})--(\ref{etwentyfive})
have to be taken, which introduces an
ambiguity in the resulting coupling constants.
evaluated. The experimental
errors in the branching fractions are somewhat large
but the masses of the particles involved are
relatively well known. We have used the standard
formulae \cite{PD}

\begin{eqnarray}
\label{etwentysix}
\widehat{f} & = & f(\widehat{x})+{1 \over 2}
\sigma^2 f''(\widehat{x})
\end{eqnarray}

\begin{eqnarray}
\label{etwentyseven}
\sigma_{f} & = & \sigma |f'(\widehat{x})|
\end{eqnarray}

for $f(x)=\sqrt{x}$ ($x=R_{\gamma}^0$ or $R_{\gamma}^+$),
where $\widehat{x}$, $\widehat{f}$ and $\sigma$,
$\sigma_{f}$ are the mean values and standard
deviations of the respective distributions.
Using expressions (\ref{etwentysix})--(\ref{etwentyseven})
instead of their linearized versions is
important due to the large experimental error in
$R_{\gamma}^+$. From that data we then find

\begin{eqnarray}
\label{etwentyeight}
|{\lambda' \over g}+{2 \over 3}{\lambda \over g}| & = &
(0.863 \pm 0.075) GeV^{-1}
\end{eqnarray}

and

\begin{eqnarray}
\label{etwentynine}
|{\lambda' \over g}-{1 \over 3}{\lambda \over g}| & = &
(0.089 \pm 0.178) GeV^{-1}
\end{eqnarray}

The two errors in eq.(\ref{etwentyeight})--
(\ref{etwentynine}) can in principle
be correlated. The main sources of correlations are the
experimental efficiencies in the detection of the $\pi^0$
and $\gamma$. Since the contributions of the efficiencies
to the net errors are small \cite{expr}, possible
correlations were neglected and the errors in the
determination of single $\lambda/g$ and $\lambda'/g$
were combined in quadrature. There are four solutions
for the ratios $\lambda/g$ and $\lambda'/g$:
$a)$ $\lambda/|g|=k(0.77 \pm 0.19)$GeV$^{-1}$,
$\lambda'/|g|=k(0.35 \pm 0.12)$GeV$^{-1}$ and
$b)$ $\lambda/|g|=k(0.95 \pm 0.19)$GeV$^{-1}$,
$\lambda'/|g|=k(0.23 \pm 0.12)$GeV$^{-1}$,
where $k$ can be $+1$ or $-1$. For the individual determination
of $\lambda$, $\lambda'$ and $g$, one has to fix
one parameter. Choosing $|\lambda|=(0.60 \pm 0.11)$
GeV$^{-1}$ as determined in \cite{G3} minimized the
errors in these quantities. We
present in Table \ref{tab1} the four possible solutions
of (\ref{etwentyeight})--(\ref{etwentynine}) for
$\lambda$ and $|g|$, together with the
combinations $\lambda' +2 \lambda /3$ and
$\lambda' - \lambda/3$, which will be used in our
calculations in section 5.

{\footnotesize
\begin{table}[h]
\begin{center}
\begin{tabular}{|c|c|c|c|c|c|}\hline
\raisebox{-2mm}  & {\rule[2mm]{0mm}{2mm} $\lambda$ [GeV$^{-1}$]}
& $\lambda'$ [GeV$^{-1}$] &
$|g|$ & $(\lambda' + {2 \over 3}\lambda)$ [GeV$^{-1}$] &
$(\lambda' - {1 \over 3}\lambda)$ [GeV$^{-1}$] \\ \hline
$1$ & $+0.60 \pm 0.11$ & $+0.29 \pm 0.12$ &
$0.82 \pm 0.20$ & $+0.71 \pm 0.18$ & $+0.07 \pm 0.15$\\
$2$ & $-0.60 \pm 0.11$ & $-0.29 \pm 0.12$ &
$0.82 \pm 0.20$ & $-0.71 \pm 0.18$ & $-0.07 \pm 0.15$\\
$3$ & $+0.60 \pm 0.11$ & $+0.15 \pm 0.09$ &
$0.66 \pm 0.13$ & $+0.57 \pm 0.12$ & $-0.06 \pm 0.12$\\
$4$ & $-0.60 \pm 0.11$ & $-0.15 \pm 0.09$ &
$0.66 \pm 0.13$ & $-0.57 \pm 0.12$ & $+0.06 \pm 0.12$\\ \hline
\end{tabular}
\caption{Four possible solutions from (28)-(29) with $\lambda$
determined by [13].}
\label{tab1}
\end{center}
\end{table}
}

The experimental value
$|g|=0.57 \pm 0.13$ (see for example \cite{G3}
and references therein for a discussion of these
experimental values) and the approximate validity
of the equation $|\lambda'| \simeq 1/(6 m_c)$
with $m_c \simeq 1.5$ GeV slightly favour solutions
$3)$ and $4)$.

Our approach is different from \cite{CG}, \cite{ABJ} and
\cite{LLY}, since we do not use any quark model
prediction for the parameter $\lambda'$ but treat it
on an equal footing with the parameter $\lambda$, so that both
are considered as purely phenomenological. Nevertheless
we were able to obtain reasonably good precision in
the determination of model parameters.

\vspace{.5cm}
\newpage
\begin{center}
{\bf 4. Weak Lagrangian for light and heavy mesons}
\end{center}

In addition to strong and electromagnetic
interactions, we must also address the
weak decays within these scheme. We will follow
the approach of \cite{G2,MW} and use an effective
current between the heavy mesons and the light mesons.
The weak current is ${L_Q}_{a}^{\mu} = {\bar q}_{a}
\gamma^{\mu}(1- \gamma_{5}) Q$ and it transforms
as $({\bar 3}_{L}, 1_{R})$. The lowest dimension
operator with the same transformation properties
but with meson degrees of freedom is

\begin{eqnarray}
\label{ethirty}
{J_Q}_{a}^{\mu} & = & \frac{1}{2} i \alpha Tr [\gamma^{\mu}
(1 - \gamma_{5})H_{b}u_{ba}^{\dag}]\\
& + & \alpha_{1}  Tr [\gamma_{5} H_{b} (\hat \rho^{\mu}
- {\cal V}^{\mu})_{bc} u_{ca}^{\dag}] + \cdot \cdot \cdot
\nonumber
\end{eqnarray}

where the ellipsis denote terms vanishing
in the limit $m_{q} \rightarrow 0$, $m_{Q}
\rightarrow \infty$ or terms with derivatives.

The light mesons decay constants $f_{P,V}$
are defined by the usual relations

\begin{eqnarray}
\label{ethirtyone}
<0|{J_q}_{\mu}^{a b}(0)|P_i (p)> & = &
i f_P {\lambda_i^{a b} \over \sqrt{2}} p_{\mu}\nonumber\\
<0|{J_q}_{\mu}^{a b}(0)|V_i (\epsilon,p)> & = &
f_V {\lambda_i^{a b} \over \sqrt{2}} m_V \epsilon_{\mu}
\end{eqnarray}

Here $\lambda_i^{a b}$, $i=1,\ldots,8$, $a,b=1,2,3$ are
the usual eight Gell-Mann $3 \times 3$ matrices, normalized
as $Tr(\lambda_i \lambda_j)=2 \delta_{i j}$. In the chiral
limit $m_q \to 0$ the above decay constants are related
through $f_P=f_V/\sqrt{a}=f$ for all $P$, $V$.
Similarly we can define the heavy meson decay
constants by \cite{MW},

\begin{eqnarray}
\label{ethirtytwo}
<0|{J_Q}_{\mu}^{a}(0)|D^b (p)> & = &
-i f_D \delta^{a b} m_D v_{\mu}\nonumber\\
<0|{J_Q}_{\mu}^{a}(0)| D^{* b} (\epsilon,p)> & = &
i f_{D^*} \delta^{a b} m_{D^*} \epsilon_{\mu}
\end{eqnarray}

In the heavy quark limit $m_Q \to \infty$ we have
$m_D=m_{D^*} \to \infty$ and $f_D=f_{D^*}$.
The constant $\alpha$   in eq. (\ref{ethirty}) can be fixed
in this limit by taking the matrix elements of
$J_{a}^{\mu}$ between the heavy meson
state and the vacuum, with the result

\begin{eqnarray}
\label{ethirtythree}
\alpha & = & f_{P} {\sqrt m_{P}}
\end{eqnarray}

Unfortunately, up to now, there does not exist either
a theoretical prediction or experimental data for
the other parameter, $\alpha_1$, in the current
(\ref{ethirty}).

The situation is different and better in the light
sector, where a well known prescription exists,
for deriving the weak current directly from the strong
Lagrangian \cite{buras}. Since the quark weak current
${L_q}_{a b}^{\mu} = {\bar q}_{b}
\gamma^{\mu}(1- \gamma_{5}) q_{a}$ must transform
as $({\bar 3}_{L},3_{R})$ the light
meson weak current with these transformation
properties can be obtained from the
Lagrangian (\ref{esixteen}). Of course, one has to properly
define the covariant derivative for pseudoscalars,
enlarging the gauge group to include
the $W$-boson contributions \cite{BKY2}.
The resulting light meson part of the weak current is

\begin{eqnarray}
\label{ethirtyfour}
{J_q}^{\mu} & = & i f^2 u [{\cal A}^{\mu}+
a ( {\cal V}^{\mu}-{\hat \rho}^{\mu}) ] u^{\dag}
\end{eqnarray}

The part of the weak Lagrangian for the pseudoscalar and
vector, light and heavy mesons, which we will use, can be
written as \cite{KXC,WSB1,WSB2}

\begin{eqnarray}
\label{ethirtyfive}
{\cal L}_{W}^{eff}(\Delta c = \Delta s = 1) =
-{G \over \sqrt{2}} V_{ud} V_{cs}^{*}
[ & a_{1} ({\bar u}d)_{V-A}^\mu ({\bar s}c )_{V-A,\mu} + &
\nonumber\\
& a_{2} ({\bar s}d)_{V-A}^\mu ({\bar u}c )_{V-A,\mu} & ]
\end{eqnarray}

where $V_{ud}$, etc. are the relevant $CKM$
mixing paremeters, while $a_{1}$ and $a_{2}$ are the
QCD Wilson coefficients, which depend on a scale
$\mu$. One expects the scale to be the heavy quark mass
and we take $\mu \simeq 1.5$ GeV which
gives $a_{1} = 1.2$ and  $a_{2} = -.5$,
with an approximate  $20\%$ error.
In the factorization model
the quark currents are approximated by the
corresponding meson currents defined in
eqs. (\ref{ethirty}) and (\ref{ethirtyfour}):

\begin{eqnarray}
\label{ethirtysix}
({\bar q}_a Q)_{V-A}^\mu & \equiv &
{\bar q}_a \gamma^\mu (1-\gamma^5) Q \simeq
{J_Q}_{a}^\mu
\end{eqnarray}

\begin{eqnarray}
\label{ethirtyseven}
({\bar q}_b q_a)_{V-A}^\mu & \equiv &
{\bar q}_b \gamma^\mu (1-\gamma^5) q_a \simeq
{J_q}_{a b}^\mu
\end{eqnarray}

Many heavy meson weak nonleptonic
amplitudes \cite{KXC,WSB1,WSB2,G4} have been
calculated using the factorization approximation.
It has been shown in \cite{KXC}, however,
that for some of the $D$ meson decays there are rather
important final state interactions and the factorization
approximation can be improved by the inclusion of
the $SU(3)$ symmetry breaking effects \cite{LC}.

The authors of ref. \cite{WSB1}  have classified
the weak nonleptonic decays into three classes:
decays determined by $a_{1}$ only (class I),
decays determined by $a_{2}$ only (class II)
and decays where $a_{1}$ and $a_{2}$ amplitudes
interfere (class III). Factorization can
therefore be tested in several ways. There are
the following
two cathegories of decays: "quark decays",
in which the heavy quark decays while remaining
antiquark acts as a  spectator, and "annihilation
processes" in
which heavy and light quarks annihilate and
two new quarks are created. For the annihilation
processes the factorization approximation is usually
only a small contribution \cite{WSB1,WSB2,KXC}.

We are forced to use this approximation in our
calculations, since there are no better approaches
developed so far for nonleptonic weak decays.

\vspace{.5cm}

\begin{center}
{\bf 5. $D\rightarrow V \gamma$ decays}
\end{center}

The simplest radiative decays of D mesons are
into a light meson and a photon. Since the
process $D \to P \gamma$ (P is a light pseudoscalar)
is forbidden due to the requirement of gauge
invariance and chiral symmetry \cite{Ecker},
as well as angular momentum conservation,
we will concentrate on the $D\rightarrow
V \gamma$ (V is a light vector meson) decays.
Although there are not yet any data on such
processes we can predict the partial widths
and branching ratios.
We consider the only two processes which are
possible at tree level and are not Cabibbo supressed,
namely $D^{0}\rightarrow {\bar{K^{*0}}} \gamma$ and
$D^{s+}\rightarrow \rho^+ \gamma$. Both processes
have contributions from the odd-parity interaction
Lagrangian. The second one has, in addition, a direct
emission term, due to
the charged initial and final mesons.

Regarding the anomalous term, there
are two contributions which are
important. The photon can first be emitted
from the $D^0$ ($D^{s+}$) meson, which becomes
a $D^{* 0}$ ($D^{*s+}$) and then
$D^{* 0}$ ($D^{*s+}$) decays weakly into
$\bar{K^{* 0}}$ ($\rho^+$).
The other contribution comes from the weak decay of
$D^0$ ($D^{s+}$) first into an off-shell
$\bar{K^0}$ ($\pi^+$), which then
decays into $\bar{K^{*0}} \gamma$
($\rho^+ \gamma$). Both contributions
are proportional to $a_{2}$ ($a_{1}$)
(\ref{ethirtyfive}).
For the description of this amplitude
we need the $D^{*} D \gamma$ and
$K^{*} K \gamma$ ($\rho \pi \gamma$) couplings
and these couplings
were obtained in the previous section.

The amplitude for the $D^0 \rightarrow
{\bar K^{*0}} \gamma$ is

\begin{eqnarray}
\label{ethirtyeight}
A(D^0(p_{D^0}) \rightarrow {\bar K^{*0}}(p_{K^{*0}})
\gamma(q))& = & e \frac{G}{{\sqrt 2}} V_{ud}
V_{cs}^{*} a_{2} \\
& & [C^{(1)}_{D^0 K^{*0}\gamma}
\epsilon_{\mu \nu \alpha \beta}
q^{\mu} \epsilon_{\gamma}^{\nu *} v^{\alpha}
\epsilon_{K^{*0}}^{\beta *}] \nonumber
\end{eqnarray}

where

\begin{eqnarray}
\label{ethirtynine}
C^{(1)}_{D^0 K^{*0}\gamma} & = &
(C_{VV\Pi} \frac{1}{g_{V}}+ C_{V\Pi \gamma})
\frac{f_{D^0}f_{K^0} m_{D^0}^2}
{(m_{D^0}^{2} - m_{K^0}^{2})}
\frac{8 {\sqrt 2} m_{D^0}}{3 f}\nonumber\\
& + & 4 ({\lambda}^\prime + \frac{2}{3} \lambda)
f_{D^{*0}}f_{K^{*0}}
\frac{m_{D^{*0}}m_{K^{*0}}}{(m_{D^{*0}}^{2} - m_{K^{*0}}^{2})}
\sqrt{m_{D^{0}} m_{D^{*0}}}
\end{eqnarray}

Similarly, the amplitude for $D^{s+} \rightarrow
\rho^+ \gamma$ is

\begin{eqnarray}
\label{eforty}
A(D^{s+}(p_{D^s}) \rightarrow \rho^{+}(p_{\rho})
\gamma(q))& = & e \frac{G}{{\sqrt 2}} V_{ud}
V_{cs}^{*} a_{1} \nonumber \\
& & [C^{(1)}_{D^s \rho \gamma}
\epsilon_{\mu \nu \alpha \beta}
q^{\mu} \epsilon_{\gamma}^{\nu *} v^{\alpha}
\epsilon_{\rho}^{\beta *} \\
& & + i C^{(2)}_{D^s \rho \gamma}
m_{\rho} (\epsilon_\gamma^* . \epsilon_\rho^* -
\frac{\epsilon_\gamma^* . p_\rho
\epsilon_\rho^* . q}{p_\rho . q})]\nonumber
\end{eqnarray}

where

\begin{eqnarray}
\label{efortyone}
C^{(1)}_{D^s \rho \gamma} & = &
-(C_{VV\Pi} \frac{1}{g_{V}}+ C_{V\Pi \gamma})
\frac{f_{D^s}f_\pi m_{D^s}^2}
{(m_{D^s}^{2} - m_\pi^{2})}
\frac{4 {\sqrt 2} m_{D^s}}{3 f}\nonumber\\
& + & 4 (\lambda' - \frac{1}{3} \lambda)
f_{D^{*s}}f_\rho
\frac{m_{D^{*s}}m_\rho}{(m_{D^{*s}}^2 - m_\rho^2)}
\sqrt{m_{D^{s}} m_{D^{*s}}}
\end{eqnarray}

and

\begin{eqnarray}
\label{efortytwo}
C^{(2)}_{D^s \rho \gamma} & = & f_{D^s} f_\rho
\end{eqnarray}

In our numerical calculations we used the following
numerical values $C_{VV\Pi} = .423$,
$C_{V\Pi\gamma} = -3,26.10^{-2}$,
\cite{BOS,FSO}, $g_{V} = 5.8$ \cite{G2},
$f \simeq f_{\pi} = 132$ MeV, and
the other decay constants $f_{P,V}$
were taken from \cite{G4}.
It is straightforward to calculate the decay widths.
The result, of course, depends on which
numerical value we take for $(\lambda' +
2 \lambda /3)$ and $(\lambda'-\lambda/3)$.
The numerical predictions for the
decay widths and branching ratios are given in
Table \ref{tab2}, where all the four possible
choices of paramaters (see Table
\ref{tab1}) are considered.

\begin{table}[h]
\begin{center}
\begin{tabular}{|c|c|c|c|c|}\hline
\raisebox{-4mm}{Solution} & \multicolumn{2}{|c|}
{$D^{0} \to {\bar K^{*0}} \gamma$} &
\multicolumn{2}{|c|}{\rule[-2mm]{0mm}{7mm}
$D^{s+} \to \rho^+ \gamma$} \\ \cline{2-5}
& $\Gamma$ [$10^{-7}$eV] & BR [$10^{-4}$] &
$\Gamma$ [$10^{-7}$eV] & BR [$10^{-4}$] \\ \hline
$1$ & $7.8 \pm 2.5$ & $4.9 \pm 1.6$ & $4.9 \pm 2.0$ & $3.3 \pm 1.3$ \\
$2$ & $0.8 \pm 0.7$ & $0.5 \pm 0.4$ & $7.6 \pm 3.7$ & $5.1 \pm 2.5$ \\
$3$ & $5.9 \pm 1.4$ & $3.7 \pm 0.9$ & $7.0 \pm 2.9$ & $4.7 \pm 1.9$ \\
$4$ & $0.3 \pm 0.3$ & $0.2 \pm 0.2$ & $4.7 \pm 1.7$ & $3.2 \pm 1.1$ \\
\hline
\end{tabular}
\caption{Different possible predictions for the
$D^0 \to {\bar K^{* 0}} \gamma$ and
$D^{s+} \to \rho^+ \gamma$ decays. The errors are due to
uncertanties in the determination of the combinations
$(\lambda'+2 \lambda /3)$ and $(\lambda'-\lambda/3)$.}
\label{tab2}
\end{center}
\end{table}

An interesting feature can be seen from Table
\ref{tab2}: A not very precise
measurement of the $D^0 \to \bar{K^{*0}} \gamma$
decay rate is sufficient to differentiate between solutions
1)-3) and solutions 2)-4), which are predicted
to be of one order of magnitude different.
Unfortunately, due to the much larger branching ratio
for the weak decay $D^0 \to \bar{K^{*0}} \pi^0$ and
the difficulty in differentiating the photon
from the $\pi^0$ in this energy range, the
decay $D^0 \to \bar{K^*} \gamma$
has not been seen by the ARGUS collaboration
\cite{TP}. The situation is similar
for the detection of $D^{s+} \to \rho^+ \gamma$
by the ARGUS collaboration: due to the very
small branching ratio, low detector's acceptance
($3 \gamma$ events have to be measured) and
poor mass resolution, the ARGUS
data are not likely to find this decay.
But, hopefully, some experimental signals
for these radiative decays will
come from the CLEO data. The experimental
measurement of this branching ratio will determine
the relative sign between the first
and the second contributions, giving in such a way
new information on the parameters $\lambda$, $\lambda'$
and $g$.

\vskip 0.5cm
In conclusion, we used both chiral symmetry
and heavy quark symmetry to obtain an
effective strong, EM and weak
Lagrangian for the description of both light and heavy
pseudoscalar and vector mesons. In this framework we
reanalyized the $D^*$ strong and radiative decays,
obtaining without any reference to quark models,
a good determination of some of the parameters in
the effective Lagrangian.
Within the same framework and with these values for
the parameters we calculated the $D \to V \gamma$
decay widths, providing numerical predictions.
These results can be used to test the validity of the
approximations that were made in the context of
the heavy quark effective theory. At the least,
our numerical results are reasonable estimates
and provide some guidance.
In the framework developed here other D meson
radiative non-leptonic decays
($D \to VP \gamma$ or $PP \gamma$) can also be calculated
\cite{bfo}, giving estimates for future
experiments and further tests of the applicability of
HQET.

\vskip 0.5cm
{\it Acknowledgement.} This work was supported in part by the
Ministry of Science and Technology of the Republic
of Slovenia (B.B. and S.F.), the German Ministerium
f\"{u}r Forschung und Technologie under contract
No. 06 TM 732 (S.F.) and by the U.S. Department
of Energy, Division of High Energy Physics,
under grant No. DE-FG02-91-ER4086 (R.J.O.).

\vskip 0.5cm
{\it Note added.} After completing of this work
the preprint of P. Jain, A. Momen and J. Schechter
(hep-ph/9406338), where some of these results have
been independently obtained, came to our attention.
Their approach is quite similar to ours in section 3,
but they fix the parameter $\lambda'$ to be
proportional to $1/m_c$, while
we leave it free. Also we noticed a preprint by
H.Y. Cheng, C.Y. Cheung, G.L. Lin, Y.C. Lin,
T.M. Yan and H.L. Yu (hep-ph/9407303). In this preprint
the $D \rightarrow K^* \gamma$ decay rate was estimated
using an effective electromagnetic and weak
Lagrangian developed from quark diagrams
for $b \bar{d} \rightarrow c \bar{u} \gamma$.
They then made the replacement
$b \rightarrow c$ and $c \rightarrow s$.
Their result is comparable with our second
case in Table \ref{tab2}.


\vskip 0.5cm
\begin{thebibliography} {99}
\bibitem{CG} P. Cho and H. Georgi,
Phys. Lett. {\bf B296} (1992) 408.
\bibitem{ABJ} J. Amundson, C.G. Boyd,
E. Jenkins, M. Luke, A. Manohar,
J. Rosner, M. Savage and M. Wise,
Phys. Lett. {\bf B296} (1992) 415.
\bibitem{IW} N. Isgur and M. Wise,
Phys. Lett.{\bf B232} (1989) 113;
{\bf B232} (1990) 527.
\bibitem{BD} G. Burdman and J. Donoghue,
Phys. Lett.{\bf B280} (1992) 287;
\bibitem{YCC} T.M. Yan, H.Y. Cheng,
C.Y. Che
{\bf D46} (1992) 1148.
\bibitem{LLY}  H.Y. Cheng, C.Y. Cheung,
G.L. Lin, Y.C. Lin, T.M. Yan and H.L. Yu, Phys. Rev.
{\bf D47} (1993) 1030; {\bf D49} (1993) 2490.
\bibitem{PC} P. Cho, Nucl. Phys. {\bf B396} (1993) 183.
\bibitem{HG1} H. Georgi, Nucl. Phys. {\bf B348} (1991) 293.
\bibitem{HG2} H. Georgi, Nucl. Phys. {\bf B240} (1990) 447.
\bibitem{FGL} A. Falk, B. Grinstein and M.Luke, Nucl. Phys.
{\bf B357} (1991) 185.
\bibitem{G1} R. Casalbuoni, A. Deandra, N.Di Bartolomeo, R. Gatto,
F. Feruglio, G. Nardulli, Phys. Lett. {\bf B294} (1992) 106.
\bibitem{G2} R. Casalbuoni, A. Deandra, N.Di. Bartolomeo, R. Gatto,
F. Feruglio, G. Nardulli, Phys. Lett. {\bf B292} (1992) 371.
\bibitem{G3} R. Casalbuoni, A. Deandra, N.Di Bartolomeo, R. Gatto,
F. Feruglio, G. Nardulli, Phys. Lett. {\bf B299} (1993) 139.
\bibitem{MW} M. Wise, Phys. Rev. {\bf D45} (1992) 1148.
\bibitem{BKY} M. Bando, T. Kugo, S. Uehara,
K. Yamawaki and T. Yanagida, Phys. Rev. Lett.
{\bf 54} (1985) 1215;
M. Bando, T. Kugo, and K. Yamawaki,
Nucl. Phys {\bf B259} (1985) 493;
Phys. Rep. 164 (1988) 217.
\bibitem{WZ}  J. Wess and B. Zumino,
Phys. Lett. {\bf B37} (1971).
\bibitem{EW} E. Witten, Nucl. Phys. {\bf223} (1983) 422.
\bibitem{PD} Particle Data 1992,
Phys. Rev. {\bf D 44} Suppl. 1992.
\bibitem{expr} CLEO Collab., F.Butler at al.,
Phys. Rev. Lett. {\bf 69} (1992) 2041.
\bibitem{buras} W.A. Bardeen, A.J. Buras and J.-M. G\'{e}rard,
Phys. Lett. {\bf B192} (1987) 138.
\bibitem{BKY2} M. Bando, T. Kugo and K. Yamawaki,
Prog. Theor. Phys. {\bf 73} (1985) 1541.
\bibitem{KXC} A.N. Kamal, Q.P. Xu and A.Czarnecki,
Phys. Rev. {\bf D49}(1994) 1330.
\bibitem{WSB1} M.Bauer, B. Stech and M. Wirbel,
Z. Phys. {\bf C34} (1987)103.
\bibitem{WSB2} M.Bauer, B. Stech and M. Wirbel,
Z. Phys. {\bf C34} (1987)103.
\bibitem{G4} A. Deandra, N. Di Bartolomeo,
R. Gatto and G. Nardulli,Phys. Lett. {\bf B318} (1993) 549.
\bibitem{LC} L.L.Chau, H.Y. Cheng,
preprint ITP-B-93-49 and UCD-93-31,(1994).
\bibitem{Ecker} G. Ecker, A. Pinch and E. de Rafael,
Nucl. Phys. {\bf B291}(1987) 692.
\bibitem{BOS} E. Braaten, R.J. Oakes and Sze-Man Tse,
Int. Jour. Mod . Phys.{\bf A5} (1990) 2737.
\bibitem{FSO} S. Fajfer, K. Suruliz and R.J. Oakes,
Phys. Rev. {\bf D46}(1992) 1195.
\bibitem{TP} T. Podobnik (ARGUS collaboration), private communication.
\bibitem{bfo} B. Bajc, S. Fajfer and R.J. Oakes, in preparation.
\end{thebibliography}
\end{document}